\documentclass[%
 preprint,
superscriptaddress,
 amsmath,amssymb,
 aps,
 pra,
]{revtex4-2}

\usepackage{graphicx}
\usepackage{dcolumn}
\usepackage{bm}
\usepackage{mathtools}
\usepackage{amsthm}
\usepackage{physics}
\usepackage{hyperref}

\begin{document}

\title[Noise-based learning]{Noise-based Local Learning using Stochastic Magnetic Tunnel Junctions}



\author{Kees Koenders}%
\affiliation{%
Department of Machine Learning and Neural Computing, 
 Donders Institute for Brain, Cognition and Behaviour, Radboud University, Nijmegen, the Netherlands
}%

\author{Leo Schnitzpan}%
\affiliation{%
Institute of Physics, Johannes Gutenberg University Mainz, 55099 Mainz, Germany
}%

\author{Fabian Kammerbauer}%
\affiliation{%
Institute of Physics, Johannes Gutenberg University Mainz, 55099 Mainz, Germany
}%

\author{Sinan Shu}%
\affiliation{%
Institute of Physics, Johannes Gutenberg University Mainz, 55099 Mainz, Germany
}%

\author{Gerhard Jakob}%
\affiliation{%
Institute of Physics, Johannes Gutenberg University Mainz, 55099 Mainz, Germany
}%

\author{Mathias Kl\"aui}%
\affiliation{%
Institute of Physics, Johannes Gutenberg University Mainz, 55099 Mainz, Germany
}%
\affiliation{Center for Quantum Spintronics, Norwegian University of Science and Technology, 7491 Trondheim, Norway
}%

\author{Johan H. Mentink}%
\affiliation{%
 Department of Physics, Radboud University, Nijmegen, the Netherlands
}

\author{Nasir Ahmad}%
\affiliation{%
Department of Machine Learning and Neural Computing, 
 Donders Institute for Brain, Cognition and Behaviour, Radboud University, Nijmegen, the Netherlands
}%

\author{Marcel van Gerven}%
\email{marcel.vangerven@donders.ru.nl}
\affiliation{%
Department of Machine Learning and Neural Computing, 
 Donders Institute for Brain, Cognition and Behaviour, Radboud University, Nijmegen, the Netherlands
}%

\begin{abstract}
    Brain-inspired learning in physical hardware has enormous potential to learn fast at minimal energy expenditure. One of the characteristics of biological learning systems is their ability to learn in the presence of various noise sources. Inspired by this observation, we introduce a novel noise-based learning approach for physical systems implementing  multi-layer neural networks. Simulation results show that our approach allows for effective learning whose performance approaches that of the conventional effective yet energy-costly backpropagation algorithm. Using a spintronics hardware implementation, we demonstrate experimentally that learning can  be achieved in a small network composed of physical stochastic magnetic tunnel junctions. These results provide a path towards efficient learning in general physical systems which embraces rather than mitigates the noise inherent in physical devices.
\end{abstract}


\maketitle

\section{Introduction}

Realizing learning in hardware by harnessing the physical properties of neuromorphic devices is currently a very active field of research owing to the potential to realize much more energy-efficient and conceptually new routes towards intelligent machinery, which can adapt to its environment at minimal energy expenditure~\cite{Kaspar2021,Markovic2020}. 

In recent years, various unconventional learning algorithms have been proposed as putative approaches for physics-based learning~\cite{scellier2017,Stern2021,Nakajima2022,lopez2023,Momeni2023,doremaele2024}. Contrastive methods such as equilibrium propagation (EP) have raised particular interest because they circumvent the expensive backpropagation (BP) of loss gradients. In EP, learning arises by comparing the steady state of a system run in a free phase and weakly clamped phase~\cite{scellier2017,Kendall2020,Stern2021}. 

A key challenge of these kinds of approaches is its sensitivity to the underlying device variability. For example, exact twins are required for EP \cite{Dillavou2022}.
Although progress has been made to remove this stringent constraint~\cite{Laborieux2022,Anisetti2023}, it is well-known that learning in biological systems can proceed by embracing noise as a principle for learning~\cite{Faisal2008,Shimizu2021}. This is very interesting as well for physical hardware, in which noise is inherent to physical devices and emerges naturally and prominently when scaling to small spatial dimensions. 

Purely stochastic hardware has recently attracted great interest with the development of probabilistic bits (p-bits)~\cite{Camsari2017,Camsari2019,Grimaldi2023,Si2024}. However, learning in stochastic networks is much more challenging. For example, direct implementations of Boltzmann-like contrastive learning rules require the evaluation of the spatial correlations between node outputs, which is feasible for small networks but does not scale well to large and dense networks~\cite{Kaiser2022}. To circumvent this problem, a local learning algorithm that can directly leverage the noise of a single node alone would be desirable. Moreover, for many important generative models, measurement of correlations between node outputs is not sufficient for learning, which strongly limits potential hardware applications.

Here we focus on a different route to learning which, instead of avoiding noise or mitigating its impact, directly harnesses the intrinsic device noise for learning. To this end, we focus on the recently developed decorrelated activity-based node perturbation (DANP) algorithm~\cite{Dalm2023}. We show that with this algorithm it is feasible to learn by direct injection of physical noise at individual nodes. The DANP algorithm goes beyond the classical node perturbation algorithm~\cite{dembo1990model,cauwenberghs1992fast} and, in contrast to the classical formulation, has excellent convergence properties. The advantage of our approach is that learning requires just two forward passes in a physical network under different noise perturbations, thereby bypassing the need for measuring the system in a clean (unperturbed) phase. Moreover, our algorithm does not require access to the noise fluctuations or their spatial correlations themselves and the error signal can be easily extracted from a differential measurement of node activities, which is straightforward to realize in practice.


The demonstration of noise-based local learning is generally feasible with any neuromorphic material featuring stochastic behavior, such as memristive devices~\cite{Lanza22}. Here, we have judiciously chosen to work with magnetic systems, which are inherently well suited for unconventional computing due to intrinsic non-linearity and tunable non-volatility~\cite{Grollier20, Markovic2020}. In particular, magnetic systems feature specific advantages for stochastic hardware since the stochastic behavior stems from the spin degrees of freedom and hence does not feature physical motion of atoms. This leads to superior endurance as compared to other memristive materials. For example, the stochastic thermally induced dynamics can be exploited in magnetic systems for probabilistic computing \cite{Zazvorka2019} and Brownian reservoir computing~\cite{Raab2022}. Most conspicuous have been stochastic magnetic tunnel junctions (sMTJs)~\cite{Camsari2019}. In such systems, the random switching can be tuned over a wide range of frequencies and biasing the switching is easily feasible by coupling to magnetic layers, external fields or using spin torques induced by injected currents~\cite{Kanai2021}.

In this work, we extend the applicability of sMTJs to noise-based learning and explicitly harness the experimentally obtained hardware-generated noise of stochastic magnetic tunnel junctions (sMTJs) for node perturbation. In particular, we show that direct injection of physical noise from an sMTJ to the nodes of the network suffices to realize learning. This is a significant advance over previous work, which demonstrated DANP-based learning only using idealized physically-unrealistic Gaussian noise. In contrast, in our work, the actual sMTJ noise is directly harnessed as a resource for learning, which potentially can lead to scalable and robust learning in physical systems.

\section{Methods}

\subsection{Noise-based learning}

\subsubsection*{The learning problem}

The learning problem is set up as follows. We consider a fully-connected neural network consisting of $L$ layers (nonlinear transformations), each of which contains $N_l$ neurons. 
Let ${x}_{0}$ be the network input.
The output of the $l$-th layer is given by
\begin{equation*}
    {x}_{l} = f\left({a}_l\right) 
\end{equation*}
where ${a}_l = {W}_l {x}_{l-1}$ is the pre-activation with
weight matrix 
${W}_l$, $f$ is the activation function and ${x}_{l}$ is the output of layer $l=1, \dots, L$.
The loss is defined as
$$
\mathcal{L} = ||y^* - y||^2 
$$
with $y^*$ the target output and $y = x_L$ the predicted output. 
To minimize the loss, we consider learning rules which update the weights of such a network using update steps
\begin{equation*}
{W}_l \gets {W}_l - \eta \Delta {W}_l 
\end{equation*}
where $\eta$ is a small, constant learning rate, and $\Delta {W}_l$ is a parameter update. 
The regular gradient descent update is given by
\begin{equation}
\var{W}_l^\textrm{GD} 
= \left\langle \frac{\partial \mathcal{L}}{\partial {W_l}} \right\rangle 
\end{equation}
where the expectation is estimated from the empirical distribution consisting of $M$ input-output pairs.

\subsubsection*{Node perturbation and forward gradients}

Node perturbation (NP) is an alternative to gradient descent, which does not require global transmission of exact gradients through the network. NP relies on comparing the network states of a clean pass and a noise-perturbed pass, in which noise is injected into the pre-activation of each layer to yield a perturbed output
\begin{equation}\label{eq:npfwdpass}
{\tilde{x}}_{l} = f\left({\tilde{a}}_l + {\epsilon}_l\right) \,,
\end{equation}
where ${\epsilon}_l \sim {\mathcal{N}}({0}, \sigma^2 {I}_l)$ is a noise vector and ${I}_l$ is an $N_l \times N_l$ identity matrix.
We define the loss differential as
$\var{\mathcal{L}} = \mathcal{L}({\tilde{x}}_L) - \mathcal{L}({x}_L)$ with $\mathcal{L}({\tilde{x}}_L)$ the loss for the noisy output and $\mathcal{L}({x}_L)$ the loss for the clean output. 
Conventional node perturbation employs the update rule
\begin{equation}    \var{W}_l^\textrm{NP} = \sigma^{-2} \left\langle \var{\mathcal{L}} {\epsilon}_l {x}_{l-1}^\top \right\rangle \,.
\end{equation}
The issue with this formulation is that it does not perform well relative to gradient descent, requires a clean forward pass, and assumes that the noise is measurable.

In~\cite{Dalm2023}, activity-based node perturbation (ANP) was introduced as an alternative to regular NP whose updates are more aligned with the gradient direction, similar in spirit to the forward gradient approach of Ref.~\cite{Baydin2022, Ren2022}. ANP employs the more advanced update rule
\begin{equation}
\var{W}_l^\textrm{ANP} = N \left\langle \: \var{\mathcal{L}} \:  \frac{\var{\alpha_l}}{||\var{{\alpha}}||^2} \: {x}_{l-1}^\top \right\rangle \,,
\end{equation}
where $N = \sum_{l=1}^L N_l$ is the total number of units in the network, $\var{\alpha}_l = {\tilde{a}}^{(1)}_l - \tilde{{a}}^{(2)}_l$ is the activity difference at layer $l$ for two forward passes perturbed by different noise realisations, and $\var{a} = (\var{a_1},\ldots,\var{a_L})$ is the concatenation of all activity differences.

Note that ANP does not require a clean forward pass and instead can compare two noisy passes without the need for access to the noise signal. This is ideally suited for application in noisy physical hardware, which can provide controlled noise as detailed below. 

\subsubsection*{Input decorrelation}
For single-layer networks, we can show that ANP is robust, highly performant, and trains extremely well.
However, although ANP is more aligned with the true gradient, still performance lags significantly behind that of gradient descent in multi-layer networks. It was shown in Ref.~\cite{Dalm2023} that this performance gap can be overcome by ensuring that the inputs to each network layer are decorrelated.
That is, using a decorrelating transform $\bar{x}_l = R_l x_l$, we replace $x_l$ in the preceding by its decorrelated representation $\bar{x}_l$ for multi-layer networks.
It was shown in~\cite{Ahmad2022} that the decorrelation matrix $R_l$ can be learned in an iterative manner using the update rule
\begin{equation*}
{R}_l \gets {R}_l - \epsilon \left\langle \bar{x}_l  \bar{x}_l^\top - \textrm{diag}\left( \bar{x}_l^2 \right)  \right\rangle{R}_l  
\end{equation*}
where $\epsilon$ is a small constant learning rate and ${R}_l$ is initialised as the identity matrix. Integrating this into the preceding leads to decorrelated activity-based node perturbation (DANP), which has been shown to approach the performance of stochastic gradient descent in multi-layer networks~\cite{Dalm2023}.

\subsection{Stochastic magnetic tunnel junctions}
Our goal is to demonstrate how noise-based learning can be implemented in networks comprised of stochastic magnetic tunnel junctions. SMTJs are magnetoresistive elements, where the thermal fluctuation of the magnetization leads to random fluctuations of the resistance~\cite{Schnitzspan2023, Camsari2019}. The origin of the resistance fluctuations is the fluctuation of a ferromagnetic free layer in a magnetic tunnel junction while the fixed reference layer is designed to be stable over the timescale of the device lifetime.  By tuning the anisotropies, one can in particular obtain two-level fluctuations leading to a volatile resistive element with telegraph noise~\cite{Hayakawa2021}. If the energy barrier between local energy minima is of the order of $k_B T$ the switching occurs on easily measurable timescales. This volatile behavior of the magnetic free layer is called superparamagnetism and for an MTJ with a typically designed uniaxial anisotropy, this results in a two-level switching with dwell times down to nanoseconds~\cite{Safranski2021, Schnitzspan2023}.

The fluctuation time, respectively dwell time $\tau$ can be described in the ``macrospin'' approximation by the Néel-Arrhenius law~\cite{Neel1949}:
\begin{equation}
    \tau = \tau_0 \exp(E_b / k_B T) 
\end{equation}
where $E_b$ is the energy barrier between the states, $T$ is the temperature, and $\tau_0$ is the attempt time.

Here, the magnetic tunnel junction is based on a CoFeB/MgO/CoFeB interface exhibiting a tunnel magnetoresistance ratio of over 100\,\% and a resistance area product of approximately 15\,$\Omega$µm$^2$.
The tunnel magnetoresistance (TMR) stack used for the MTJ devices was deposited by RF- and DC-magnetron sputtering (Singulus Rotaris) at room temperature and subsequently annealed at 300\,$^{\circ}$C for 1\,h at 300\,mT applied in-plane field. 
The composition of the stack, derived from a previously developed configuration~\cite{Schnitzspan2020}, is as follows, with film thickness specified in nanometers:
Ta(10)/Ru(10)/Ta(10)/PtMn(20)/CoFe(2.2)/Ru(0.8)/ CoFeB(2.4)/MgO(1.1)/CoFeB(3.0)/Ta(10)/Ru(30). MTJ nanopillars were patterned in circular shapes of the size of 60\,nm by means of electron beam lithography.
The MTJs exhibit a low in-plane uniaxial anisotropy at the ferromagnetic free layer, resulting in superparamagnetic switching between the parallel (low resistance) and antiparallel (high resistance) state.\\

Due to their inherent randomness, full CMOS compatibility, robustness, and energy efficiency, MTJ devices are excellent candidates for neuromorphic hardware harnessing noise-based learning algorithms like ANP.
This noise source not only allows for the generation of correlation-free true randomness~\cite{Schnitzspan2023}, but also for the generation of noise based on specific probability distributions through appropriate interconnection of multiple sMTJs~\cite{Schnitzspan2023b}.
An additional advantage lies in the tunability of the MTJ state probability and fluctuation rate through applied magnetic fields or injected currents exerting torques on the free layer.

%

\subsection{Learning with sMTJ noise}

\subsubsection*{Analog learning}
\begin{figure}
    \begin{minipage}[]{0.49\textwidth}
    \includegraphics[width=0.98\linewidth]{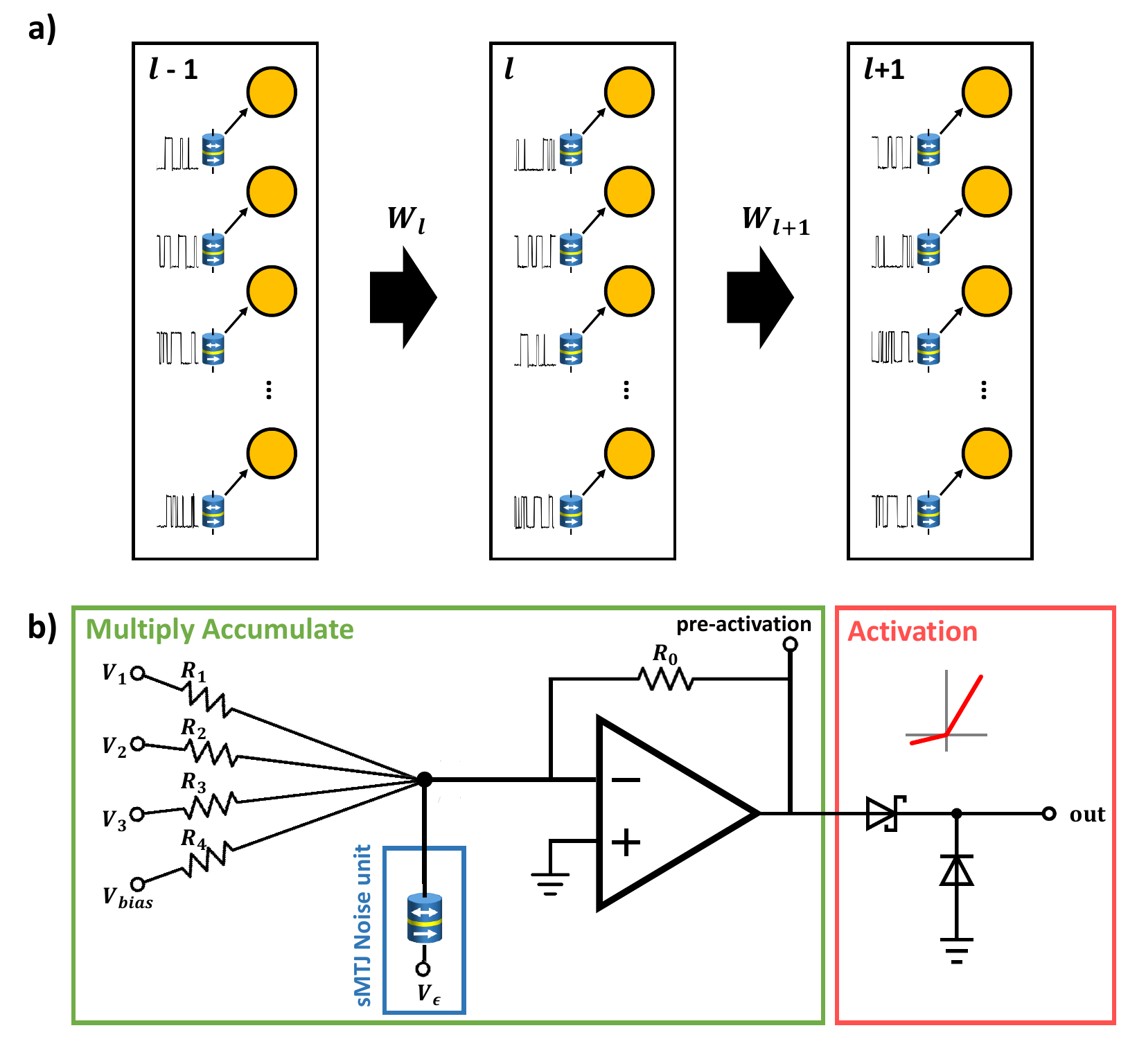}
    \end{minipage}%
    \begin{minipage}[]{0.49\textwidth}
    \caption{a) Illustration of the ANP network implementation with superparamagnetic MTJs as a perturbation source. A single sMTJ provides the required random noise for each node at each layer in the network.
    b) ANP-perceptron circuit as an analog implementation of sMTJs and CMOS transistors. A sMTJ is used as a random stochastic perturbation source in the circuit. Weighting is implemented as (analog) resistive weights and the ReLU-like activation function is implemented by two diodes. The multiply accumulate (MAC) operation is carried out by an operational amplifier in the configuration of a summing amplifier.}
    \label{fig:NP_network_sketch_and_perceptron}
    \end{minipage}
\end{figure}
Since the main advantages of noise-based learning with sMTJs are realised in fully analog circuitry,
we provide a minimal ANP-perceptron circuit design illustrated in Figure~\ref{fig:NP_network_sketch_and_perceptron}.
Voltages function as inputs while resistances function as weights.
Noise is injected into the node as a volatile voltage source, due to the fluctuating resistance of the sMTJ, to an operational amplifier in a voltage adder circuit, 
which enables the multiply-accumulate operation of the node.
A leaky ReLU activation function is implemented through a voltage divider circuit with two diodes.
The design can be expanded upon by connecting multiple sMTJs in series to deliver multi-level noise. This is the case for a resistance-based setup, as shown here.
Alternatively, one might want to work with conductance in which case a parallel setup should be used.  
A single sMTJ will provide a two-level noise and a series of $n$ sMTJs will provide multilevel noise ranging from $n+1$, 
if the tunnel magnetoresistance (TMR) of each sMTJ is exactly the same, to $2^n$ if the TMR of each sMTJ is distinct.
This offers more fine-grained perturbations, which can improve overall training performance.
Additionally, by adding more sMTJs the average fluctuation rate is increased as this is given by $\Gamma= 1/\tau^*= \sum_i \tau_i^{-1}$, where $\tau^*$ is the average dwell time of the sMTJ ensemble and $\tau_i$ the average dwell time of the $i$-th sMTJ.
A higher fluctuation rate increases the noise generation bandwidth,
allowing for shorter iteration times.
A complete design including CMOS circuitry which implements the learning dynamics,
while feasible due to the local nature of the learning rule,
is reserved for future work.

\subsubsection*{Measured sMTJ noise}
In our experiments, to demonstrate learning with real-world sMTJ noise,
we use a time series measured from a single sMTJ as perturbation.
This is done in a linear fashion where neuron $i+1$ receives the signal sampled at the next timestep compared to neuron $i$.
There will be large autocorrelations by construction with this approach, and thus this serves as a worst-case scenario where a single sMTJ provides noise for an entire network.
\\
In addition to using real sMTJ noise in larger networks,
we provide a proof-of-concept where an sMTJ is measured in real-time by an Arduino to perform online learning.

\subsubsection*{Simulated sMTJ noise}
To show the potential of large-scale application of sMTJ's, we employ an approach where we construct a hidden Markov model with two states and a noisy observation.
The parameters of this model are computed statistically from the single sMTJ time series.
Details can be found in  Appendix~\ref{app:hyperparameters}.

\subsubsection*{Approximated sMTJ noise}
A single sMTJ can be considered as a Bernoulli random variable with dwell times distributed according to a Poisson distribution.
For this approximation we assume that when noise is needed, the dwell time will already have passed.
Thus we draw from a symmetric Bernoulli distribution,
scaling the result by a factor $\alpha$.
If multi-level noise is desired, $h$ samples are drawn and summed.


\section{Results}

\subsection{Noise characterisation}
\begin{figure}
    \begin{minipage}{0.5\textwidth}
    \includegraphics[width=0.92\linewidth]{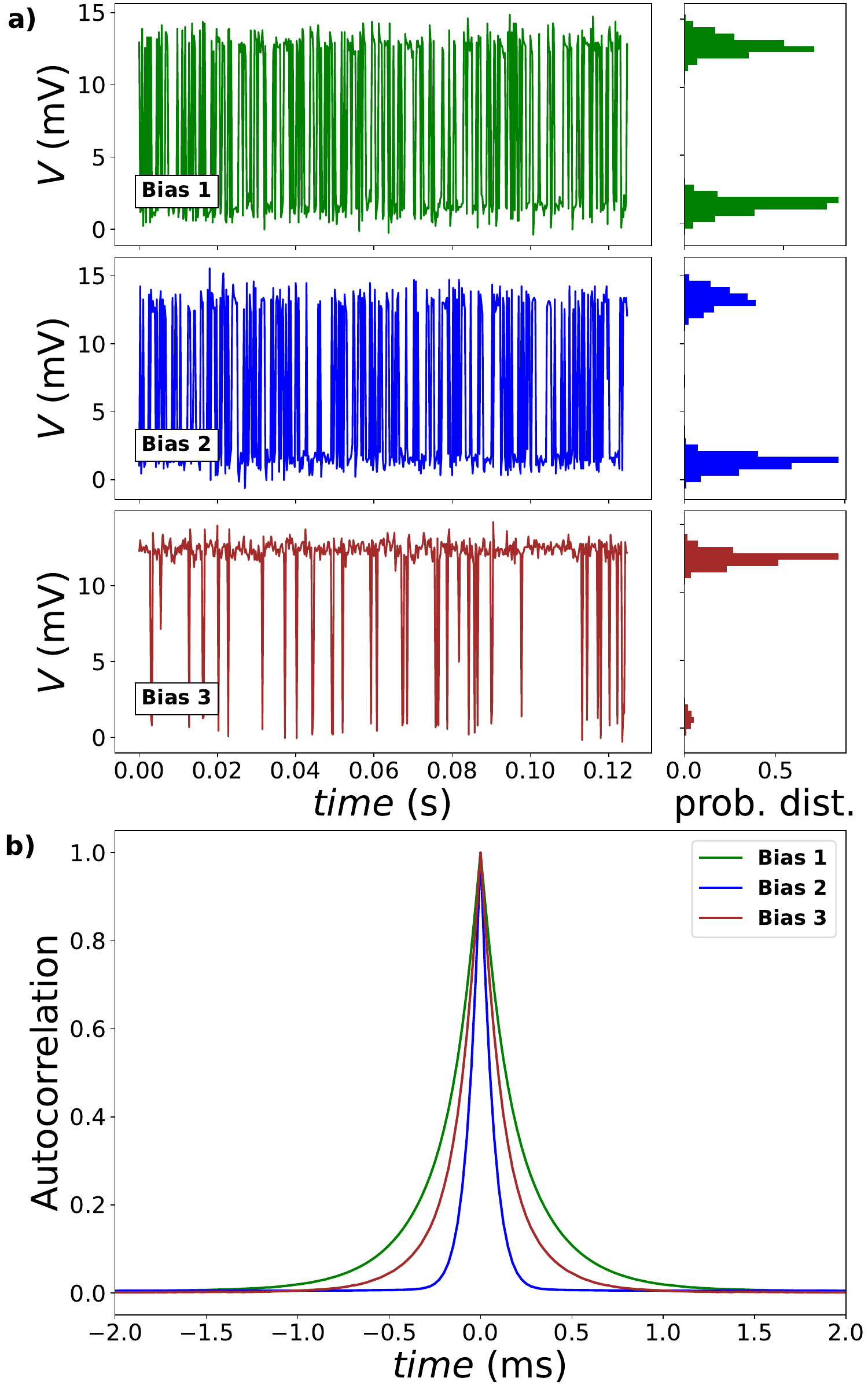}
    \end{minipage}~
    \begin{minipage}{0.5\textwidth}
    \caption{a) Time series of a sMTJ circuit for the generation of noise under three different biases originating from external in-plane magnetic fields ranging from approximately -3 to +3 mT. Top, center and bottom display the switching behaviour of the same sMTJ, along with histograms illustrating their respective probability distributions.
    b) Autocorrelation of the sMTJ noise signal for sMTJs with the three different field biases shown in a). The autocorrelation function was calculated from 2500\,s long time series measurements.}
    \label{fig:SMTJ_time_series}
    \end{minipage}
\end{figure}
In Figure \ref{fig:SMTJ_time_series} the stochastic random noise generated by superparamagnetic tunnel junctions is illustrated.
The data was acquired for 2500\,s by an oscilloscope (sampling rate of 40\,kHz adapted to the dwell time of the sample used).
A measurement of a single stochastic MTJ is shown in Figure \ref{fig:SMTJ_time_series}a. The state probability is random but can be tuned electrically via spin-transfer torques~\cite{Schnitzspan2023} or magnetically via external in-plane magnetic fields.
Biases may also arise intrinsically due to adjacent magnetic fields or as a result of the sMTJ fabrication process.
We generate two-level noise with different probability distributions via external field biases. 


When considering the autocorrelation function of these time series, the correlation time scale of the noise signal can be obtained, which specifically depends on the individual fluctuation rates of each sMTJ.
The autocorrelation for the three different measured time series of Figure~\ref{fig:SMTJ_time_series}a is plotted in Figure~\ref{fig:SMTJ_time_series}b.
Depending on the sMTJ bias, the signal is decorrelated after a millisecond at the latest in the sample used. However, we note that with the demonstrated nanosecond dwell times, this can be orders of magnitude faster.

\subsection{Training single-layer sMTJ networks}

\begin{figure*}[!ht]
\centering
  \includegraphics[width=0.92\textwidth]{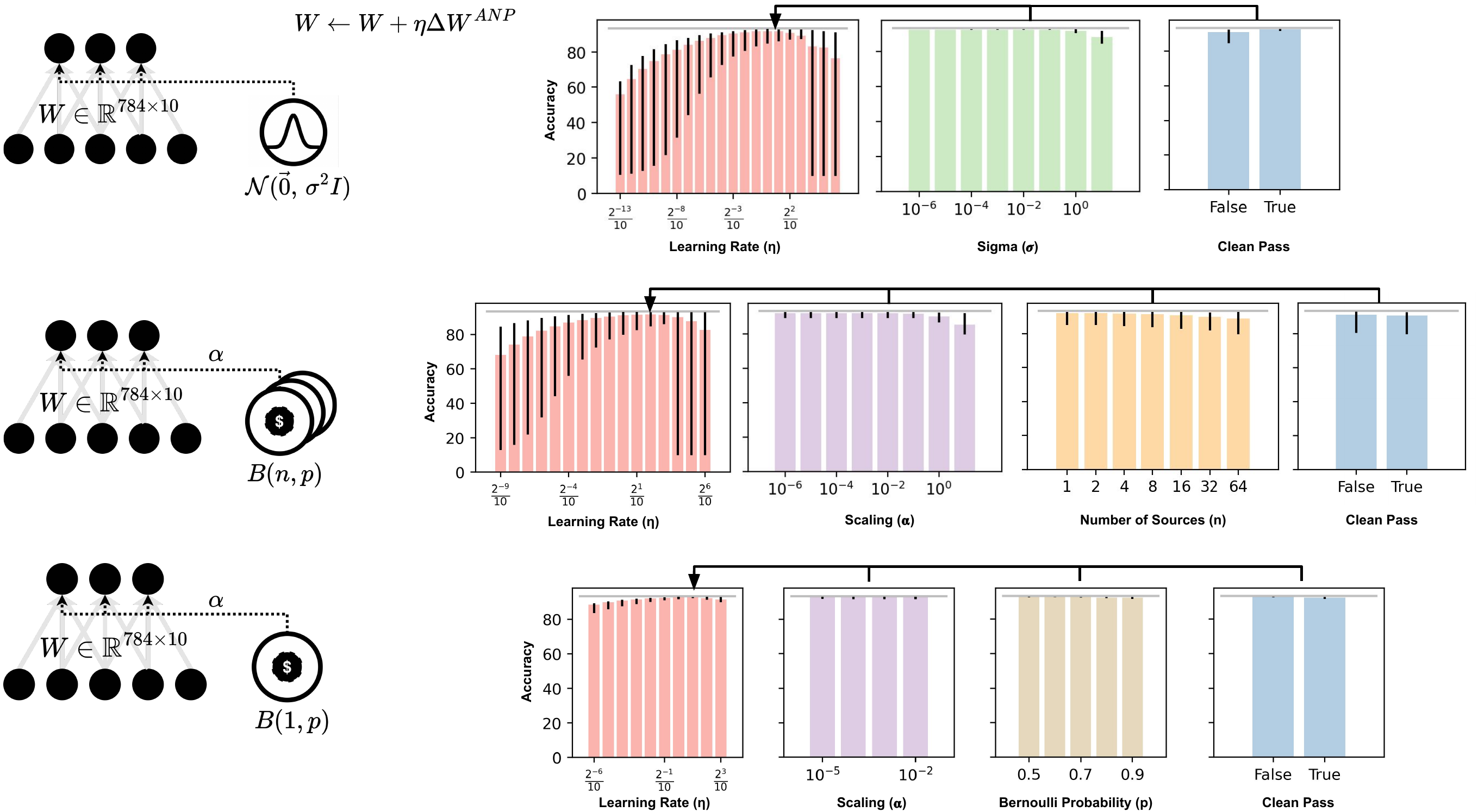}
  \caption{Parameter robustness of ANP in a single linear-layer network trained on the MNIST dataset. The accuracies shown are the training accuracies reached after 100 epochs of training with the selected parameters with a categorical cross-entropy loss. Note that these results show slices of a parameter sweep hypercube, such that error bars show maximum and minimum performance for a given parameter value as measured across all options for the \textit{other} remaining hyperparameters. Thus error bars can be relatively high if they are susceptible to change in other hyperparameters. Hyperparameters include, the learning rate ($\eta$), noise variance ($\sigma^2$), a multiplicative scaling factor applied to noise ($\alpha$), Bernoulli noise source number (n) and Bernoulli probability (p). Gray horizontal line indicates the best BP performance based upon the best simulation outcome when BP networks are trained for 100 epochs over learning rates $\eta \in \{0.1, 0.1 \cdot 2^{-1}, \ldots, 0.1 \cdot 2^{-9}\}$.}
\label{fig:onelayer}
\end{figure*}

We tested the validity of our approach by examining learning behaviour in simulated sMTJ networks.
Figure~\ref{fig:onelayer} shows the final performance of single-layer networks (trained for 100 epochs) using the ANP formulation.

Importantly, the ANP approach is robust to a significant range of choices in parameters and noise sources.
Figure~\ref{fig:onelayer} shows how training of a single layer is robust to learning rate and noise scale, whether one has Gaussian or Binomial distribution sampled noise.
Furthermore, in the regime of the proposed sMTJ setup (Bernoulli distribution) we can also observe successful learning.
Note that the gray horizontal lines here correspond to the maximum accuracy of a BP-trained network after 100 epochs when trained across the same set of learning rates.

\begin{figure*}[!ht]
  \includegraphics[width=0.98\textwidth]{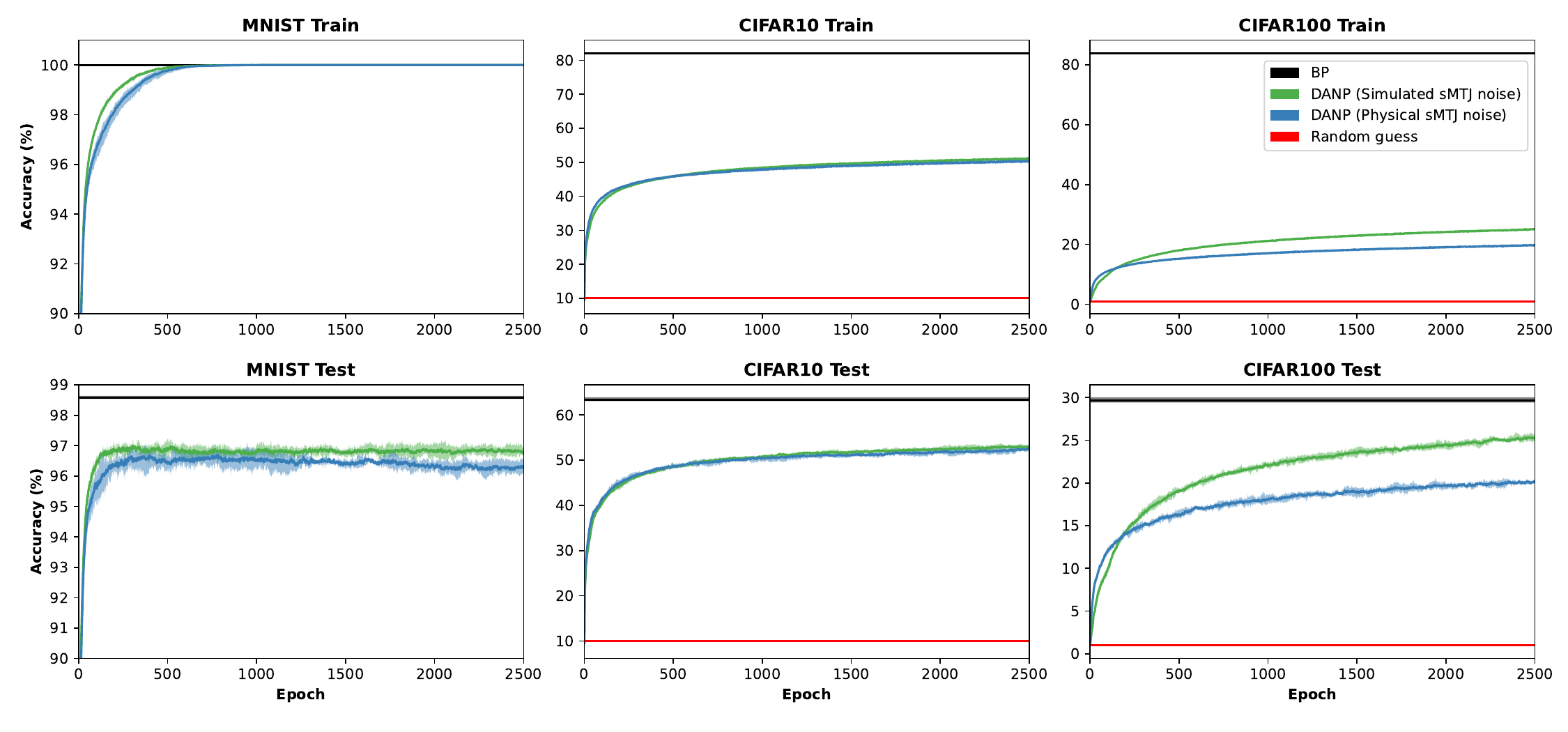}
  \caption{Training convergence of multi-layer neural networks trained by BP vs.\ DANP with noise either simulated from multiple sMTJs or sampled from a real single sMTJ on three different tasks: MNIST, CIFAR10 and CIFAR100. For CIFAR, training samples have been augmented by randomly cropping and flipping them. These networks consist of the input layer, followed by three hidden layers of 500 units, and finally by an output layer of 10 or 100 units. Training parameters were chosen based upon a parameter sweep, see Appendix \ref{app:hyperparameters}. Maximum accuracy achieved by BP is shown as a theoretical upper bound as a black line, while the random guess lower bound is shown in red. Envelopes represent the maximum and minimum accuracies across five randomly seeded repeats. Noise for the DANP simulations was either 1) simulated with an independent sMTJ noise source simulated for every node of a network, or 2) was taken from collected data of a single sMTJ, such that the time series data collected for the sMTJ was directly used as noise for nodes in a serial fashion. That is, each node is given noise in order from the sMTJ time series from the input layer of the network to the output layer. In all cases, the DANP algorithm operates with two noisy passes.}
\label{fig:multilayer}
\end{figure*}
\subsection{Training multi-layer sMTJ networks}

Next, we wished to determine whether multilayer neural networks can be trained with real-world sampled noise.
We therefore applied our approach for training in multilayer perceptrons with noise sampled from a long recording session with a single sMTJ.

Figure~\ref{fig:multilayer} shows the results of these simulations in which fully connected networks (comprised of an input layer, three hidden layers of 500 nodes, and finally an output layer of 10 or 100 nodes) are trained by the DANP algorithm with noise sampled directly from a time series of sMTJ-generated noise.
For this training, no clean passes were computed and instead only noisy passes were computed for the DANP rule.
Since backward gradient machinery is not feasible to implement in physical learning systems,
we include only maximum values for BP-trained variants of these networks as a theoretical optimum.

In these results, we can see that the DANP trained simulations are capable of learning to a significant degree.
Their performance does not reach as high as that of BP, but still perform significantly above the 10\% and 1\% random guess accuracy on the test set for CIFAR10 and CIFAR100, respectively. Particularly, for CIFAR100, test performance gets close to that of BP and is still increasing at the end of the simulation. For both CIFAR results, DANP has slightly higher test than train accuracy. This can be attributed to the fact that the test set has no augmentations, and the tendency for (D)ANP to generalise well, limiting overfitting on the train set~\cite{fernandez2024}.

Even in the extreme case, in which we use real-world data sampled from a single sMTJ and provide this as noise to each node sequentially, learning closely follows that of the network with simulated sMTJ noise.
When injecting this measured noise into the nodes of this system, subsequent samples of noise from that single sMTJ were taken as the injected noise, meaning that the networks experienced noise with high degrees of autocorrelation between neighbouring units.
This can be seen as a worst-case scenario for learning with real-world noise, and yet networks are still shown to yield reasonable test accuracy.
Thus, the ability to train a state-of-the-art multi-layer neural network without backward passes of the conventional computationally expensive backpropagation algorithm but instead with only the injection of noise drawn from a real physical system shows promise even in the worst case scenario of a single noise source feeding all units sequentially.
Appendix~\ref{app:hyperparameters} provides the hyperparameters and tuning process for these simulations.

\subsection{Experimental validation}

\begin{figure}
\begin{minipage}{0.5\textwidth}
\includegraphics[width=\linewidth]{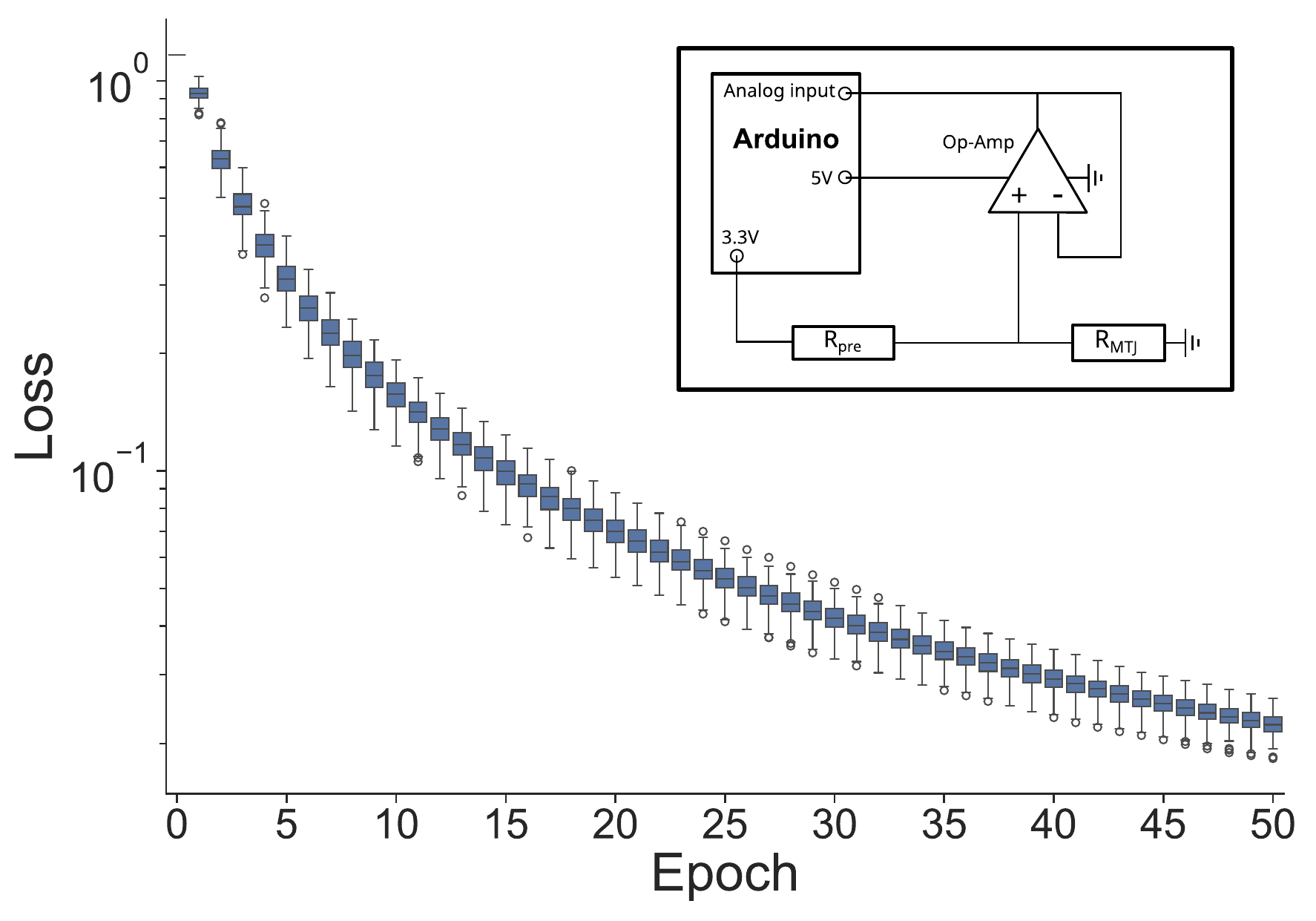}
\end{minipage}~
\begin{minipage}{0.5\textwidth}
\caption{Boxplot of the loss of a ANP network employing a single sMTJ as noise source for 100 epochs. Whiskers show extremal values excluding outliers, while the box represents 25th to 75th percentile values. A total of 100 samples were used per epoch. The inset displays the employed circuitry to read a single sMTJ via an Arduino Uno. Training parameters are reported in Appendix~\ref{app:live_params}.}
\label{fig:empirical}
\end{minipage}
\end{figure}

Finally, we demonstrate effective sMTJ learning applied within physical hardware. For this, we make use of the experimental setup shown in the inset of Figure~\ref{fig:empirical}. We employ an Arduino Uno, which has 14-bit resolution on its analog pins to read the voltage level at the sMTJ and supply voltage via 3.3 V pins. The essential part is a voltage divider between a resistor of 100 k$\Omega$ and the MTJ, whose resistance is in the order of 3-6 k$\Omega$. An MCP601 Op-Amp is used as a unity gain buffer. The Arduino is set up with Telemetrix~\cite{TelemetrixArduino}, which allows for direct communication via Python. The network runs within Python and reads the noise directly sampled by the Arduino. 
The noise is hereby integrated via a noise-sampler function which saves a buffer array of 200 noise values, with new values continuously added and old ones removed. Upon a noise sample call, the code waits for 1 ms, and returns a shuffled list of the 200 stored values. The wait time is chosen to mimic the readout speed of the Arduino of 1 kHz, so that upon each call at least one new value is present.

To quantify learning performance, we tested the setup for its ability to approximate the input-output mapping of a ground-truth, randomly initialized, $2\times 2\times 2 \times 1$ network. A total of 100 samples were generated with the ground truth network. Figure~\ref{fig:empirical} demonstrates effective learning as weight changes reduce the loss, which is robust over 100 epochs with small deviations for the loss after 50 epochs of training.
With this, we showcase a simple and in principle scalable approach to read sMTJs and use a physical system as a source of on-the-fly generated noise in our proof-of-concept hardware device.

\section{Discussion}

Our results demonstrate both in simulations and an actual physical experiment that learning based on physical noise is a viable strategy. We show that these results are robust under variations of normal and Bernoulli distributed  physical noise. Also see~\citep{Belouze2022} for a motivation on the use of the latter.
Our findings suggest that our strategy may be generalizable to many physical noise sources, opening the door towards noise-based learning in a wide range of physical systems.



A next step in the development of physical learning machines would be to incorporate not only the noise process but also the forward propagation of activity and input decorrelation, as well as the updating of parameters in the physical substrate. 
Importantly, activity-based node perturbation relies on local operations only, save for a global feedback signal that encodes and scales the loss difference. Additionally, the iterative decorrelation is local at the layer-level.
These properties offer enormous potential for learning in physical hardware, reducing the circuitry required leading to significant energy savings.

Beyond the current line of research, we also see potential for the further integration of noise-based learning approaches into noisy physical materials and noisy models.
These include formulating noise-based learning within Hamiltonian system descriptions of physical systems, within sampling based models for parallel sampling as well as learning, and much more.

With the development of our advanced noise-based learning approach, we contribute to the development of a new generation of intelligent systems, directly realizable in a range of (noisy, analog) physics-based hardware implementations~\cite{Jaeger2023}.

\section{Acknowledgements}
We acknowledge support by the EU, in particular by the Horizon 2020 Framework Program of the European Commission under FETOpen
grant agreement no. 863155 (s-Nebula) and ERC-2019-SyG no. 856538 (3D MAGiC) and the Horizon Europe project no. 101070290
(NIMFEIA). The group in Mainz acknowledges support by the Deutsche Forschungsgemeinschaft (DFG, German Research Foundation) projects 403502522 (SPP 2137 Skyrmionics), 49741853, and 268565370 (SFB TRR173 projects A01, A12 and B02) as well as TopDyn and
the Zeiss foundation through the Center for Emergent Algorithmic Intelligence. L.S. was supported by the Max Planck Graduate Center
with the Johannes Gutenberg-Universität Mainz (MPGC).
J.H.M. acknowledges funding from the VIDI project no. 223.157 (CHASEMAG) which is financed by the Dutch Research Council (NWO).
M.K. thanks the Research Council of Norway for support through its Centres of Excellence funding scheme, project number 262633 ‘QuSpin’. 
M.G. and N.A. are supported by the Dutch Brain Interface Initiative (DBI2) with project number 024.005.022 of the research programme Gravitation, which is financed by the Dutch Ministry of Education, Culture and Science (OCW) via the Dutch Research Council (NWO).










\clearpage

\appendix
\section{Simulation-based hyperparameters}\label{app:hyperparameters}

To produce the simulations shown in Figure~\ref{fig:multilayer}, multi-layer neural networks were constructed using PyTorch. These consist of three fully-connected hidden layers of 500 units following the input and finally a readout layer of either 10 or 100 units. The Adam optimizer was used for both BP and DANP, with $\beta_1=0.9$, $\beta_2=0.999$, $\epsilon=10^{-8}$.
A categorical cross entropy loss was used for training.

\begin{table}[!ht]
    \caption{Hyperparameters corresponding to simulations shown in in Figure~\ref{fig:multilayer}.}
    \centering
    \begin{tabular}{l|l|l|l}
       Learning algorithm & Dataset  & $\eta$ & $\epsilon$ \\
         \hline
         BP & MNIST & $0.1 \cdot 2^{-9}$ & - \\
         BP & CIFAR10 & $0.1\cdot 2^{-14}$ & - \\
         BP & CIFAR100 & $0.1\cdot 2^{-13}$ & - \\
         DANP -- simulated noise&MNIST & $0.1\cdot 2^{-11}$ & $0.1\cdot 2^{-19}$\\
         DANP -- simulated noise&CIFAR10 & $0.1\cdot 2^{-12}$ & $0.1\cdot 2^{-19}$ \\
         DANP -- simulated noise&CIFAR100 & $0.1\cdot 2^{-12}$ & $0.1\cdot 2^{-20}$\\ 
         DANP -- measured noise&MNIST & $0.1\cdot 2^{-11}$ & $0.1\cdot 2^{-19}$\\
         DANP -- measured noise&CIFAR10 & $0.1\cdot 2^{-12}$ & $0.1\cdot 2^{-19}$ \\
         DANP -- measured noise&CIFAR100 & $0.1\cdot 2^{-10}$ & $0.1\cdot 2^{-14}$ 
    \end{tabular}
    \label{tab:multilayerhyperparameters}
\end{table}
To determine optimal parameters, a parameter sweep was carried out over a range of learning rates for $\eta$ and $\epsilon$.
For every (pair of) values, a single seed of the neural network was trained for 1000 epochs to find the parameters which produced highest test accuracy at the end of training. We selected on test accuracy to mitigate overfitting.
The learning rates were tested from the set $\{0.1, 0.1 \cdot 2^{-1}, 0.1 \cdot 2^{-2}, \ldots, 0.1 \cdot 2^{-20}\}$.
The final selected learning rates are shown in Table~\ref{tab:multilayerhyperparameters}.

For the simulated sMTJ simulations, sMTJ responses were simulated as a hidden markov model (HMM) with two states and noisy observation.
The transition probability between the two HMM states was computed statistically from our single sMTJ noise data and was symmetric ($p=0.0809$), the states had two distinct mean value outputs ($\mu_1 = 0.0480$, $\mu_2 = 0.0362$) and the observation noise of these states was gaussian (mean zero and $\sigma = 0.001$).

\section{Hardware-based hyperparameters}\label{app:live_params}

To showcase the possibility of live learning using physical sMTJs we use a 2$\times$2$\times$2$\times$1 network as described in the main text. Hereby we tested a parameter sweep over the learning rates $\eta \in \{1, 5\cdot 10^{-1}, 1\cdot 10^{-1}, 7\cdot 10^{-2}, 4\cdot 10^{-2}, 1\cdot 10^{-2}, 7\cdot 10^{-3}, 4\cdot 10^{-3}, 1\cdot 10^{-3}, 5\cdot 10^{-4}\}$ and $\epsilon \in \{ 1\cdot 10^{-2}, 5\cdot 10^{-3}, 1\cdot 10^{-3}, 7\cdot 10^{-4}, 4\cdot 10^{-4}, 1\cdot 10^{-4}, 7\cdot 10^{-5}, 4\cdot 10^{-5}, 1\cdot 10^{-5}, 5\cdot 10^{-6}\}$. The finally selected parameters to produce Figure \ref{fig:empirical} are $\eta= 1 \cdot 10^{-3}$ and  $\epsilon = 1 \cdot 10^{-4}$.

\section{Code and reproducibility}
The code used to produce the results in Figure~\ref{fig:multilayer} is available on GitHub at \url{https://github.com/artcogsys/MTJ-Node-Perturbation.git}. The sMTJ time series data is available on Zenodo~\cite{sMTJ2025}.

\bibliography{bibliography}

\begin{thebibliography}{41}%
\makeatletter
\providecommand \@ifxundefined [1]{%
 \@ifx{#1\undefined}
}%
\providecommand \@ifnum [1]{%
 \ifnum #1\expandafter \@firstoftwo
 \else \expandafter \@secondoftwo
 \fi
}%
\providecommand \@ifx [1]{%
 \ifx #1\expandafter \@firstoftwo
 \else \expandafter \@secondoftwo
 \fi
}%
\providecommand \natexlab [1]{#1}%
\providecommand \enquote  [1]{``#1''}%
\providecommand \bibnamefont  [1]{#1}%
\providecommand \bibfnamefont [1]{#1}%
\providecommand \citenamefont [1]{#1}%
\providecommand \href@noop [0]{\@secondoftwo}%
\providecommand \href [0]{\begingroup \@sanitize@url \@href}%
\providecommand \@href[1]{\@@startlink{#1}\@@href}%
\providecommand \@@href[1]{\endgroup#1\@@endlink}%
\providecommand \@sanitize@url [0]{\catcode `\\12\catcode `\$12\catcode `\&12\catcode `\#12\catcode `\^12\catcode `\_12\catcode `\%12\relax}%
\providecommand \@@startlink[1]{}%
\providecommand \@@endlink[0]{}%
\providecommand \url  [0]{\begingroup\@sanitize@url \@url }%
\providecommand \@url [1]{\endgroup\@href {#1}{\urlprefix }}%
\providecommand \urlprefix  [0]{URL }%
\providecommand \Eprint [0]{\href }%
\providecommand \doibase [0]{https://doi.org/}%
\providecommand \selectlanguage [0]{\@gobble}%
\providecommand \bibinfo  [0]{\@secondoftwo}%
\providecommand \bibfield  [0]{\@secondoftwo}%
\providecommand \translation [1]{[#1]}%
\providecommand \BibitemOpen [0]{}%
\providecommand \bibitemStop [0]{}%
\providecommand \bibitemNoStop [0]{.\EOS\space}%
\providecommand \EOS [0]{\spacefactor3000\relax}%
\providecommand \BibitemShut  [1]{\csname bibitem#1\endcsname}%
\let\auto@bib@innerbib\@empty
\bibitem [{\citenamefont {Kaspar}\ \emph {et~al.}(2021)\citenamefont {Kaspar}, \citenamefont {Ravoo}, \citenamefont {van~der Wiel}, \citenamefont {Wegner},\ and\ \citenamefont {Pernice}}]{Kaspar2021}%
  \BibitemOpen
  \bibfield  {author} {\bibinfo {author} {\bibfnamefont {C.}~\bibnamefont {Kaspar}}, \bibinfo {author} {\bibfnamefont {B.~J.}\ \bibnamefont {Ravoo}}, \bibinfo {author} {\bibfnamefont {W.~G.}\ \bibnamefont {van~der Wiel}}, \bibinfo {author} {\bibfnamefont {S.~V.}\ \bibnamefont {Wegner}},\ and\ \bibinfo {author} {\bibfnamefont {W.~H.}\ \bibnamefont {Pernice}},\ }\bibfield  {title} {\bibinfo {title} {The rise of intelligent matter},\ }\href {https://doi.org/10.1038/s41586-021-03453-y} {\bibfield  {journal} {\bibinfo  {journal} {Nature}\ }\textbf {\bibinfo {volume} {594}},\ \bibinfo {pages} {345} (\bibinfo {year} {2021})}\BibitemShut {NoStop}%
\bibitem [{\citenamefont {Markovic}\ \emph {et~al.}(2020)\citenamefont {Markovic}, \citenamefont {Mizrahi}, \citenamefont {Querlioz},\ and\ \citenamefont {Grollier}}]{Markovic2020}%
  \BibitemOpen
  \bibfield  {author} {\bibinfo {author} {\bibfnamefont {D.}~\bibnamefont {Markovic}}, \bibinfo {author} {\bibfnamefont {A.}~\bibnamefont {Mizrahi}}, \bibinfo {author} {\bibfnamefont {D.}~\bibnamefont {Querlioz}},\ and\ \bibinfo {author} {\bibfnamefont {J.}~\bibnamefont {Grollier}},\ }\bibfield  {title} {\bibinfo {title} {Physics for neuromorphic computing},\ }\href {https://doi.org/0.1038/s42254-020-0208-2} {\bibfield  {journal} {\bibinfo  {journal} {Nat. Rev. Phys.}\ }\textbf {\bibinfo {volume} {2}},\ \bibinfo {pages} {499} (\bibinfo {year} {2020})}\BibitemShut {NoStop}%
\bibitem [{\citenamefont {Scellier}\ and\ \citenamefont {Bengio}(2017)}]{scellier2017}%
  \BibitemOpen
  \bibfield  {author} {\bibinfo {author} {\bibfnamefont {B.}~\bibnamefont {Scellier}}\ and\ \bibinfo {author} {\bibfnamefont {Y.}~\bibnamefont {Bengio}},\ }\bibfield  {title} {\bibinfo {title} {Equilibrium propagation: Bridging the gap between energy-based models and backpropagation},\ }\href {https://doi.org/10.3389/fncom.2017.00024} {\bibfield  {journal} {\bibinfo  {journal} {Front. comput. neurosci.}\ }\textbf {\bibinfo {volume} {11}} (\bibinfo {year} {2017})}\BibitemShut {NoStop}%
\bibitem [{\citenamefont {Stern}\ \emph {et~al.}(2021)\citenamefont {Stern}, \citenamefont {Hexner}, \citenamefont {Rocks},\ and\ \citenamefont {Liu}}]{Stern2021}%
  \BibitemOpen
  \bibfield  {author} {\bibinfo {author} {\bibfnamefont {M.}~\bibnamefont {Stern}}, \bibinfo {author} {\bibfnamefont {D.}~\bibnamefont {Hexner}}, \bibinfo {author} {\bibfnamefont {J.~W.}\ \bibnamefont {Rocks}},\ and\ \bibinfo {author} {\bibfnamefont {A.~J.}\ \bibnamefont {Liu}},\ }\bibfield  {title} {\bibinfo {title} {Supervised learning in physical networks: From machine learning to learning machines},\ }\href {https://doi.org/10.1103/PhysRevX.11.021045} {\bibfield  {journal} {\bibinfo  {journal} {Phys. Rev. X}\ }\textbf {\bibinfo {volume} {11}} (\bibinfo {year} {2021})}\BibitemShut {NoStop}%
\bibitem [{\citenamefont {Nakajima}\ \emph {et~al.}(2022)\citenamefont {Nakajima}, \citenamefont {Inoue}, \citenamefont {Tanaka}, \citenamefont {Kuniyoshi}, \citenamefont {Hashimoto},\ and\ \citenamefont {Nakajima}}]{Nakajima2022}%
  \BibitemOpen
  \bibfield  {author} {\bibinfo {author} {\bibfnamefont {M.}~\bibnamefont {Nakajima}}, \bibinfo {author} {\bibfnamefont {K.}~\bibnamefont {Inoue}}, \bibinfo {author} {\bibfnamefont {K.}~\bibnamefont {Tanaka}}, \bibinfo {author} {\bibfnamefont {Y.}~\bibnamefont {Kuniyoshi}}, \bibinfo {author} {\bibfnamefont {T.}~\bibnamefont {Hashimoto}},\ and\ \bibinfo {author} {\bibfnamefont {K.}~\bibnamefont {Nakajima}},\ }\bibfield  {title} {\bibinfo {title} {Physical deep learning with biologically inspired training method: gradient-free approach for physical hardware},\ }\href {https://doi.org/10.1038/s41467-022-35216-2} {\bibfield  {journal} {\bibinfo  {journal} {Nat. Commun.}\ }\textbf {\bibinfo {volume} {13}},\ \bibinfo {pages} {1} (\bibinfo {year} {2022})}\BibitemShut {NoStop}%
\bibitem [{\citenamefont {L\'opez-Pastor}\ and\ \citenamefont {Marquardt}(2023)}]{lopez2023}%
  \BibitemOpen
  \bibfield  {author} {\bibinfo {author} {\bibfnamefont {V.}~\bibnamefont {L\'opez-Pastor}}\ and\ \bibinfo {author} {\bibfnamefont {F.}~\bibnamefont {Marquardt}},\ }\bibfield  {title} {\bibinfo {title} {Self-learning machines based on hamiltonian echo backpropagation},\ }\href {https://doi.org/10.1103/PhysRevX.13.031020} {\bibfield  {journal} {\bibinfo  {journal} {Phys. Rev. X}\ }\textbf {\bibinfo {volume} {13}},\ \bibinfo {pages} {031020} (\bibinfo {year} {2023})}\BibitemShut {NoStop}%
\bibitem [{\citenamefont {Momeni}\ \emph {et~al.}(2023)\citenamefont {Momeni}, \citenamefont {Rahmani}, \citenamefont {Malléjac}, \citenamefont {del Hougne},\ and\ \citenamefont {Fleury}}]{Momeni2023}%
  \BibitemOpen
  \bibfield  {author} {\bibinfo {author} {\bibfnamefont {A.}~\bibnamefont {Momeni}}, \bibinfo {author} {\bibfnamefont {B.}~\bibnamefont {Rahmani}}, \bibinfo {author} {\bibfnamefont {M.}~\bibnamefont {Malléjac}}, \bibinfo {author} {\bibfnamefont {P.}~\bibnamefont {del Hougne}},\ and\ \bibinfo {author} {\bibfnamefont {R.}~\bibnamefont {Fleury}},\ }\bibfield  {title} {\bibinfo {title} {Backpropagation-free training of deep physical neural networks},\ }\href {https://doi.org/10.1126/science.adi8474} {\bibfield  {journal} {\bibinfo  {journal} {Science}\ } (\bibinfo {year} {2023})}\BibitemShut {NoStop}%
\bibitem [{\citenamefont {Van~Doremaele}\ \emph {et~al.}(2024)\citenamefont {Van~Doremaele}, \citenamefont {Stevens}, \citenamefont {Ringeling}, \citenamefont {Spolaor}, \citenamefont {Fattori},\ and\ \citenamefont {van~de Burgt}}]{doremaele2024}%
  \BibitemOpen
  \bibfield  {author} {\bibinfo {author} {\bibfnamefont {E.~R.~W.}\ \bibnamefont {Van~Doremaele}}, \bibinfo {author} {\bibfnamefont {T.}~\bibnamefont {Stevens}}, \bibinfo {author} {\bibfnamefont {S.}~\bibnamefont {Ringeling}}, \bibinfo {author} {\bibfnamefont {S.}~\bibnamefont {Spolaor}}, \bibinfo {author} {\bibfnamefont {M.}~\bibnamefont {Fattori}},\ and\ \bibinfo {author} {\bibfnamefont {Y.}~\bibnamefont {van~de Burgt}},\ }\bibfield  {title} {\bibinfo {title} {Hardware implementation of backpropagation using progressive gradient descent for in situ training of multilayer neural networks},\ }\href {https://doi.org/10.1126/sciadv.ado8999} {\bibfield  {journal} {\bibinfo  {journal} {Sci. Adv.}\ }\textbf {\bibinfo {volume} {10}},\ \bibinfo {pages} {8999} (\bibinfo {year} {2024})}\BibitemShut {NoStop}%
\bibitem [{\citenamefont {Kendall}\ \emph {et~al.}(2020)\citenamefont {Kendall}, \citenamefont {Pantone}, \citenamefont {Manickavasagam}, \citenamefont {Bengio},\ and\ \citenamefont {Scellier}}]{Kendall2020}%
  \BibitemOpen
  \bibfield  {author} {\bibinfo {author} {\bibfnamefont {J.}~\bibnamefont {Kendall}}, \bibinfo {author} {\bibfnamefont {R.}~\bibnamefont {Pantone}}, \bibinfo {author} {\bibfnamefont {K.}~\bibnamefont {Manickavasagam}}, \bibinfo {author} {\bibfnamefont {Y.}~\bibnamefont {Bengio}},\ and\ \bibinfo {author} {\bibfnamefont {B.}~\bibnamefont {Scellier}},\ }\bibfield  {title} {\bibinfo {title} {Training end-to-end analog neural networks with equilibrium propagation},\ }\href {https://doi.org/https://doi.org/10.48550/arXiv.2006.01981} {\bibfield  {journal} {\bibinfo  {journal} {arXiv:2006.01981 [cs.NE]}\ } (\bibinfo {year} {2020})}\BibitemShut {NoStop}%
\bibitem [{\citenamefont {Dillavou}\ \emph {et~al.}(2022)\citenamefont {Dillavou}, \citenamefont {Stern}, \citenamefont {Liu},\ and\ \citenamefont {Durian}}]{Dillavou2022}%
  \BibitemOpen
  \bibfield  {author} {\bibinfo {author} {\bibfnamefont {S.}~\bibnamefont {Dillavou}}, \bibinfo {author} {\bibfnamefont {M.}~\bibnamefont {Stern}}, \bibinfo {author} {\bibfnamefont {A.~J.}\ \bibnamefont {Liu}},\ and\ \bibinfo {author} {\bibfnamefont {D.~J.}\ \bibnamefont {Durian}},\ }\bibfield  {title} {\bibinfo {title} {Demonstration of decentralized physics-driven learning},\ }\href {https://doi.org/10.1103/PhysRevApplied.18.014040} {\bibfield  {journal} {\bibinfo  {journal} {Phys. Rev. Appl.}\ }\textbf {\bibinfo {volume} {18}},\ \bibinfo {pages} {014040} (\bibinfo {year} {2022})}\BibitemShut {NoStop}%
\bibitem [{\citenamefont {Laborieux}\ and\ \citenamefont {Zenke}(2022)}]{Laborieux2022}%
  \BibitemOpen
  \bibfield  {author} {\bibinfo {author} {\bibfnamefont {A.}~\bibnamefont {Laborieux}}\ and\ \bibinfo {author} {\bibfnamefont {F.}~\bibnamefont {Zenke}},\ }\bibfield  {title} {\bibinfo {title} {Holomorphic equilibrium propagation computes exact gradients through finite size oscillations},\ }in\ \href {https://proceedings.neurips.cc/paper_files/paper/2022/file/545a114e655f9d25ba0d56ea9a01fc6e-Paper-Conference.pdf} {\emph {\bibinfo {booktitle} {NeurIPS}}},\ Vol.~\bibinfo {volume} {36}\ (\bibinfo {year} {2022})\ pp.\ \bibinfo {pages} {1--14}\BibitemShut {NoStop}%
\bibitem [{\citenamefont {Anisetti}\ \emph {et~al.}(2023)\citenamefont {Anisetti}, \citenamefont {Scellier},\ and\ \citenamefont {Schwarz}}]{Anisetti2023}%
  \BibitemOpen
  \bibfield  {author} {\bibinfo {author} {\bibfnamefont {V.~R.}\ \bibnamefont {Anisetti}}, \bibinfo {author} {\bibfnamefont {B.}~\bibnamefont {Scellier}},\ and\ \bibinfo {author} {\bibfnamefont {J.~M.}\ \bibnamefont {Schwarz}},\ }\bibfield  {title} {\bibinfo {title} {Learning by non-interfering feedback chemical signaling in physical networks},\ }\href {https://doi.org/10.1103/PhysRevResearch.5.023024} {\bibfield  {journal} {\bibinfo  {journal} {Phys. Rev. Res.}\ }\textbf {\bibinfo {volume} {5}} (\bibinfo {year} {2023})}\BibitemShut {NoStop}%
\bibitem [{\citenamefont {Faisal}\ \emph {et~al.}(2008)\citenamefont {Faisal}, \citenamefont {Selen},\ and\ \citenamefont {Wolpert}}]{Faisal2008}%
  \BibitemOpen
  \bibfield  {author} {\bibinfo {author} {\bibfnamefont {A.~A.}\ \bibnamefont {Faisal}}, \bibinfo {author} {\bibfnamefont {L.~P.~J.}\ \bibnamefont {Selen}},\ and\ \bibinfo {author} {\bibfnamefont {D.~M.}\ \bibnamefont {Wolpert}},\ }\bibfield  {title} {\bibinfo {title} {Noise in the nervous system},\ }\href {https://doi.org/10.1038/nrn2258} {\bibfield  {journal} {\bibinfo  {journal} {Nat. Rev. Neurosci.}\ }\textbf {\bibinfo {volume} {9}},\ \bibinfo {pages} {292} (\bibinfo {year} {2008})}\BibitemShut {NoStop}%
\bibitem [{\citenamefont {Shimizu}\ \emph {et~al.}(2021)\citenamefont {Shimizu}, \citenamefont {Yoshida}, \citenamefont {Kasai},\ and\ \citenamefont {Toyoizumi}}]{Shimizu2021}%
  \BibitemOpen
  \bibfield  {author} {\bibinfo {author} {\bibfnamefont {G.}~\bibnamefont {Shimizu}}, \bibinfo {author} {\bibfnamefont {K.}~\bibnamefont {Yoshida}}, \bibinfo {author} {\bibfnamefont {H.}~\bibnamefont {Kasai}},\ and\ \bibinfo {author} {\bibfnamefont {T.}~\bibnamefont {Toyoizumi}},\ }\bibfield  {title} {\bibinfo {title} {Computational roles of intrinsic synaptic dynamics},\ }\href {https://doi.org/https://doi.org/10.1016/j.conb.2021.06.002} {\bibfield  {journal} {\bibinfo  {journal} {Curr. Opin. Neurobiol.}\ }\textbf {\bibinfo {volume} {70}},\ \bibinfo {pages} {34} (\bibinfo {year} {2021})}\BibitemShut {NoStop}%
\bibitem [{\citenamefont {Camsari}\ \emph {et~al.}(2017)\citenamefont {Camsari}, \citenamefont {Faria}, \citenamefont {Sutton},\ and\ \citenamefont {Datta}}]{Camsari2017}%
  \BibitemOpen
  \bibfield  {author} {\bibinfo {author} {\bibfnamefont {K.~Y.}\ \bibnamefont {Camsari}}, \bibinfo {author} {\bibfnamefont {R.}~\bibnamefont {Faria}}, \bibinfo {author} {\bibfnamefont {B.~M.}\ \bibnamefont {Sutton}},\ and\ \bibinfo {author} {\bibfnamefont {S.}~\bibnamefont {Datta}},\ }\bibfield  {title} {\bibinfo {title} {Stochastic $p$-bits for invertible logic},\ }\href {https://doi.org/10.1103/PhysRevX.7.031014} {\bibfield  {journal} {\bibinfo  {journal} {Phys. Rev. X}\ }\textbf {\bibinfo {volume} {7}},\ \bibinfo {pages} {031014} (\bibinfo {year} {2017})}\BibitemShut {NoStop}%
\bibitem [{\citenamefont {Borders}\ \emph {et~al.}(2019)\citenamefont {Borders}, \citenamefont {Pervaiz}, \citenamefont {Fukami}, \citenamefont {Camsari}, \citenamefont {Ohno},\ and\ \citenamefont {Datta}}]{Camsari2019}%
  \BibitemOpen
  \bibfield  {author} {\bibinfo {author} {\bibfnamefont {W.~A.}\ \bibnamefont {Borders}}, \bibinfo {author} {\bibfnamefont {A.~Z.}\ \bibnamefont {Pervaiz}}, \bibinfo {author} {\bibfnamefont {S.}~\bibnamefont {Fukami}}, \bibinfo {author} {\bibfnamefont {K.~Y.}\ \bibnamefont {Camsari}}, \bibinfo {author} {\bibfnamefont {H.}~\bibnamefont {Ohno}},\ and\ \bibinfo {author} {\bibfnamefont {S.}~\bibnamefont {Datta}},\ }\bibfield  {title} {\bibinfo {title} {Integer factorization using stochastic magnetic tunnel junctions},\ }\href {https://doi.org/10.1038/s41586-019-1557-9} {\bibfield  {journal} {\bibinfo  {journal} {Nature}\ }\textbf {\bibinfo {volume} {573}},\ \bibinfo {pages} {390} (\bibinfo {year} {2019})}\BibitemShut {NoStop}%
\bibitem [{\citenamefont {Grimaldi}\ \emph {et~al.}(2023)\citenamefont {Grimaldi}, \citenamefont {Mazza}, \citenamefont {Raimondo}, \citenamefont {Tullo}, \citenamefont {Rodrigues}, \citenamefont {Camsari}, \citenamefont {Crupi}, \citenamefont {Carpentieri}, \citenamefont {Puliafito},\ and\ \citenamefont {Finocchio}}]{Grimaldi2023}%
  \BibitemOpen
  \bibfield  {author} {\bibinfo {author} {\bibfnamefont {A.}~\bibnamefont {Grimaldi}}, \bibinfo {author} {\bibfnamefont {L.}~\bibnamefont {Mazza}}, \bibinfo {author} {\bibfnamefont {E.}~\bibnamefont {Raimondo}}, \bibinfo {author} {\bibfnamefont {P.}~\bibnamefont {Tullo}}, \bibinfo {author} {\bibfnamefont {D.}~\bibnamefont {Rodrigues}}, \bibinfo {author} {\bibfnamefont {K.~Y.}\ \bibnamefont {Camsari}}, \bibinfo {author} {\bibfnamefont {V.}~\bibnamefont {Crupi}}, \bibinfo {author} {\bibfnamefont {M.}~\bibnamefont {Carpentieri}}, \bibinfo {author} {\bibfnamefont {V.}~\bibnamefont {Puliafito}},\ and\ \bibinfo {author} {\bibfnamefont {G.}~\bibnamefont {Finocchio}},\ }\bibfield  {title} {\bibinfo {title} {Evaluating spintronics-compatible implementations of ising machines},\ }\href {https://doi.org/10.1103/PhysRevApplied.20.024005} {\bibfield  {journal} {\bibinfo  {journal} {Phys. Rev. Appl.}\ }\textbf {\bibinfo {volume} {20}},\ \bibinfo {pages} {024005} (\bibinfo {year} {2023})}\BibitemShut {NoStop}%
\bibitem [{\citenamefont {Si}\ \emph {et~al.}(2024)\citenamefont {Si}, \citenamefont {Yang}, \citenamefont {Cen}, \citenamefont {Chen}, \citenamefont {Huang}, \citenamefont {Yao}, \citenamefont {Kim}, \citenamefont {Cai}, \citenamefont {Yoo}, \citenamefont {Fong},\ and\ \citenamefont {Yang}}]{Si2024}%
  \BibitemOpen
  \bibfield  {author} {\bibinfo {author} {\bibfnamefont {J.}~\bibnamefont {Si}}, \bibinfo {author} {\bibfnamefont {S.}~\bibnamefont {Yang}}, \bibinfo {author} {\bibfnamefont {Y.}~\bibnamefont {Cen}}, \bibinfo {author} {\bibfnamefont {J.}~\bibnamefont {Chen}}, \bibinfo {author} {\bibfnamefont {Y.}~\bibnamefont {Huang}}, \bibinfo {author} {\bibfnamefont {Z.}~\bibnamefont {Yao}}, \bibinfo {author} {\bibfnamefont {D.~J.}\ \bibnamefont {Kim}}, \bibinfo {author} {\bibfnamefont {K.}~\bibnamefont {Cai}}, \bibinfo {author} {\bibfnamefont {J.}~\bibnamefont {Yoo}}, \bibinfo {author} {\bibfnamefont {X.}~\bibnamefont {Fong}},\ and\ \bibinfo {author} {\bibfnamefont {H.}~\bibnamefont {Yang}},\ }\bibfield  {title} {\bibinfo {title} {Energy-efficient superparamagnetic ising machine and its application to traveling salesman problems},\ }\href {https://doi.org/10.1038/s41467-024-47818-z} {\bibfield  {journal} {\bibinfo  {journal} {Nat. Commun.}\ }\textbf {\bibinfo {volume} {15}} (\bibinfo {year} {2024})}\BibitemShut {NoStop}%
\bibitem [{\citenamefont {Kaiser}\ \emph {et~al.}(2022)\citenamefont {Kaiser}, \citenamefont {Borders}, \citenamefont {Camsari}, \citenamefont {Fukami}, \citenamefont {Ohno},\ and\ \citenamefont {Datta}}]{Kaiser2022}%
  \BibitemOpen
  \bibfield  {author} {\bibinfo {author} {\bibfnamefont {J.}~\bibnamefont {Kaiser}}, \bibinfo {author} {\bibfnamefont {W.~A.}\ \bibnamefont {Borders}}, \bibinfo {author} {\bibfnamefont {K.~Y.}\ \bibnamefont {Camsari}}, \bibinfo {author} {\bibfnamefont {S.}~\bibnamefont {Fukami}}, \bibinfo {author} {\bibfnamefont {H.}~\bibnamefont {Ohno}},\ and\ \bibinfo {author} {\bibfnamefont {S.}~\bibnamefont {Datta}},\ }\bibfield  {title} {\bibinfo {title} {Hardware-aware in situ learning based on stochastic magnetic tunnel junctions},\ }\href {https://doi.org/10.1103/PhysRevApplied.17.014016} {\bibfield  {journal} {\bibinfo  {journal} {Phys. Rev. Appl.}\ }\textbf {\bibinfo {volume} {17}},\ \bibinfo {pages} {014016} (\bibinfo {year} {2022})}\BibitemShut {NoStop}%
\bibitem [{\citenamefont {Dalm}\ \emph {et~al.}(2023)\citenamefont {Dalm}, \citenamefont {van Gerven},\ and\ \citenamefont {Ahmad}}]{Dalm2023}%
  \BibitemOpen
  \bibfield  {author} {\bibinfo {author} {\bibfnamefont {S.}~\bibnamefont {Dalm}}, \bibinfo {author} {\bibfnamefont {M.}~\bibnamefont {van Gerven}},\ and\ \bibinfo {author} {\bibfnamefont {N.}~\bibnamefont {Ahmad}},\ }\bibfield  {title} {\bibinfo {title} {Effective learning with node perturbation in deep neural networks},\ }\href {https://doi.org/https://doi.org/10.48550/arXiv.2310.00965} {\bibfield  {journal} {\bibinfo  {journal} {arXiv:2310.00965 [cs.LG]}\ } (\bibinfo {year} {2023})}\BibitemShut {NoStop}%
\bibitem [{\citenamefont {Dembo}\ and\ \citenamefont {Kailath}(1990)}]{dembo1990model}%
  \BibitemOpen
  \bibfield  {author} {\bibinfo {author} {\bibfnamefont {A.}~\bibnamefont {Dembo}}\ and\ \bibinfo {author} {\bibfnamefont {T.}~\bibnamefont {Kailath}},\ }\bibfield  {title} {\bibinfo {title} {Model-free distributed learning},\ }\href {https://doi.org/10.1109/72.80205} {\bibfield  {journal} {\bibinfo  {journal} {IEEE Trans Neural Netw.}\ }\textbf {\bibinfo {volume} {1}},\ \bibinfo {pages} {58} (\bibinfo {year} {1990})}\BibitemShut {NoStop}%
\bibitem [{\citenamefont {Cauwenberghs}(1992)}]{cauwenberghs1992fast}%
  \BibitemOpen
  \bibfield  {author} {\bibinfo {author} {\bibfnamefont {G.}~\bibnamefont {Cauwenberghs}},\ }\bibfield  {title} {\bibinfo {title} {A fast stochastic error-descent algorithm for supervised learning and optimization},\ }in\ \href {https://api.semanticscholar.org/CorpusID:1964981} {\emph {\bibinfo {booktitle} {NeurIPS}}},\ Vol.~\bibinfo {volume} {5}\ (\bibinfo {year} {1992})\BibitemShut {NoStop}%
\bibitem [{\citenamefont {Lanza}\ \emph {et~al.}(2022)\citenamefont {Lanza}, \citenamefont {Sebastian}, \citenamefont {Lu}, \citenamefont {Gallo}, \citenamefont {Chang}, \citenamefont {Akinwande}, \citenamefont {Puglisi}, \citenamefont {Alshareef}, \citenamefont {Liu},\ and\ \citenamefont {Roldan}}]{Lanza22}%
  \BibitemOpen
  \bibfield  {author} {\bibinfo {author} {\bibfnamefont {M.}~\bibnamefont {Lanza}}, \bibinfo {author} {\bibfnamefont {A.}~\bibnamefont {Sebastian}}, \bibinfo {author} {\bibfnamefont {W.~D.}\ \bibnamefont {Lu}}, \bibinfo {author} {\bibfnamefont {M.~L.}\ \bibnamefont {Gallo}}, \bibinfo {author} {\bibfnamefont {M.-F.}\ \bibnamefont {Chang}}, \bibinfo {author} {\bibfnamefont {D.}~\bibnamefont {Akinwande}}, \bibinfo {author} {\bibfnamefont {F.~M.}\ \bibnamefont {Puglisi}}, \bibinfo {author} {\bibfnamefont {H.~N.}\ \bibnamefont {Alshareef}}, \bibinfo {author} {\bibfnamefont {M.}~\bibnamefont {Liu}},\ and\ \bibinfo {author} {\bibfnamefont {J.~B.}\ \bibnamefont {Roldan}},\ }\bibfield  {title} {\bibinfo {title} {Memristive technologies for data storage, computation, encryption, and radio-frequency communication},\ }\href {https://doi.org/10.1126/science.abj9979} {\bibfield  {journal} {\bibinfo  {journal} {Science}\ }\textbf {\bibinfo {volume} {376}} (\bibinfo {year} {2022})}\BibitemShut {NoStop}%
\bibitem [{\citenamefont {Grollier}\ \emph {et~al.}(2020)\citenamefont {Grollier}, \citenamefont {Querlioz}, \citenamefont {Camsari}, \citenamefont {Everschor-Sitte}, \citenamefont {Fukami},\ and\ \citenamefont {Stiles}}]{Grollier20}%
  \BibitemOpen
  \bibfield  {author} {\bibinfo {author} {\bibfnamefont {J.}~\bibnamefont {Grollier}}, \bibinfo {author} {\bibfnamefont {D.}~\bibnamefont {Querlioz}}, \bibinfo {author} {\bibfnamefont {K.}~\bibnamefont {Camsari}}, \bibinfo {author} {\bibfnamefont {K.}~\bibnamefont {Everschor-Sitte}}, \bibinfo {author} {\bibfnamefont {S.}~\bibnamefont {Fukami}},\ and\ \bibinfo {author} {\bibfnamefont {M.}~\bibnamefont {Stiles}},\ }\bibfield  {title} {\bibinfo {title} {Neuromorphic spintronics},\ }\href {https://doi.org/10.1038/s41928-019-0360-9} {\bibfield  {journal} {\bibinfo  {journal} {Nat. Electron.}\ }\textbf {\bibinfo {volume} {3}},\ \bibinfo {pages} {360} (\bibinfo {year} {2020})}\BibitemShut {NoStop}%
\bibitem [{\citenamefont {Zázvorka}\ \emph {et~al.}(2019)\citenamefont {Zázvorka}, \citenamefont {Jakobs}, \citenamefont {Heinze}, \citenamefont {Keil}, \citenamefont {Kromin}, \citenamefont {Jaiswal}, \citenamefont {Litzius}, \citenamefont {Jakob}, \citenamefont {Virnau}, \citenamefont {Pinna}, \citenamefont {Everschor-Sitte}, \citenamefont {Rózsa}, \citenamefont {Donges}, \citenamefont {Nowak},\ and\ \citenamefont {Kläui}}]{Zazvorka2019}%
  \BibitemOpen
  \bibfield  {author} {\bibinfo {author} {\bibfnamefont {J.}~\bibnamefont {Zázvorka}}, \bibinfo {author} {\bibfnamefont {F.}~\bibnamefont {Jakobs}}, \bibinfo {author} {\bibfnamefont {D.}~\bibnamefont {Heinze}}, \bibinfo {author} {\bibfnamefont {N.}~\bibnamefont {Keil}}, \bibinfo {author} {\bibfnamefont {S.}~\bibnamefont {Kromin}}, \bibinfo {author} {\bibfnamefont {S.}~\bibnamefont {Jaiswal}}, \bibinfo {author} {\bibfnamefont {K.}~\bibnamefont {Litzius}}, \bibinfo {author} {\bibfnamefont {G.}~\bibnamefont {Jakob}}, \bibinfo {author} {\bibfnamefont {P.}~\bibnamefont {Virnau}}, \bibinfo {author} {\bibfnamefont {D.}~\bibnamefont {Pinna}}, \bibinfo {author} {\bibfnamefont {K.}~\bibnamefont {Everschor-Sitte}}, \bibinfo {author} {\bibfnamefont {L.}~\bibnamefont {Rózsa}}, \bibinfo {author} {\bibfnamefont {A.}~\bibnamefont {Donges}}, \bibinfo {author} {\bibfnamefont {U.}~\bibnamefont {Nowak}},\ and\ \bibinfo {author} {\bibfnamefont {M.}~\bibnamefont {Kläui}},\ }\bibfield  {title} {\bibinfo {title} {Thermal skyrmion
  diffusion used in a reshuffler device},\ }\href {https://doi.org/10.1038/s41565-019-0436-8} {\bibfield  {journal} {\bibinfo  {journal} {Nat. Nanotechnol.}\ }\textbf {\bibinfo {volume} {14}},\ \bibinfo {pages} {658} (\bibinfo {year} {2019})}\BibitemShut {NoStop}%
\bibitem [{\citenamefont {Raab}\ \emph {et~al.}(2022)\citenamefont {Raab}, \citenamefont {Brems}, \citenamefont {Beneke}, \citenamefont {Dohi}, \citenamefont {Rothörl}, \citenamefont {Kammerbauer}, \citenamefont {Mentink},\ and\ \citenamefont {Kläui}}]{Raab2022}%
  \BibitemOpen
  \bibfield  {author} {\bibinfo {author} {\bibfnamefont {K.}~\bibnamefont {Raab}}, \bibinfo {author} {\bibfnamefont {M.~A.}\ \bibnamefont {Brems}}, \bibinfo {author} {\bibfnamefont {G.}~\bibnamefont {Beneke}}, \bibinfo {author} {\bibfnamefont {T.}~\bibnamefont {Dohi}}, \bibinfo {author} {\bibfnamefont {J.}~\bibnamefont {Rothörl}}, \bibinfo {author} {\bibfnamefont {F.}~\bibnamefont {Kammerbauer}}, \bibinfo {author} {\bibfnamefont {J.~H.}\ \bibnamefont {Mentink}},\ and\ \bibinfo {author} {\bibfnamefont {M.}~\bibnamefont {Kläui}},\ }\bibfield  {title} {\bibinfo {title} {Brownian reservoir computing realized using geometrically confined skyrmion dynamics},\ }\href {https://doi.org/10.1038/s41467-022-34309-2} {\bibfield  {journal} {\bibinfo  {journal} {Nat. Commun.}\ }\textbf {\bibinfo {volume} {13}},\ \bibinfo {pages} {6982} (\bibinfo {year} {2022})}\BibitemShut {NoStop}%
\bibitem [{\citenamefont {Kanai}\ \emph {et~al.}(2021)\citenamefont {Kanai}, \citenamefont {Hayakawa}, \citenamefont {Ohno},\ and\ \citenamefont {Fukami}}]{Kanai2021}%
  \BibitemOpen
  \bibfield  {author} {\bibinfo {author} {\bibfnamefont {S.}~\bibnamefont {Kanai}}, \bibinfo {author} {\bibfnamefont {K.}~\bibnamefont {Hayakawa}}, \bibinfo {author} {\bibfnamefont {H.}~\bibnamefont {Ohno}},\ and\ \bibinfo {author} {\bibfnamefont {S.}~\bibnamefont {Fukami}},\ }\bibfield  {title} {\bibinfo {title} {Theory of relaxation time of stochastic nanomagnets},\ }\href {https://doi.org/10.1103/PhysRevB.103.094423} {\bibfield  {journal} {\bibinfo  {journal} {Phys. Rev. B}\ }\textbf {\bibinfo {volume} {103}},\ \bibinfo {pages} {094423} (\bibinfo {year} {2021})}\BibitemShut {NoStop}%
\bibitem [{\citenamefont {Baydin}\ \emph {et~al.}(2022)\citenamefont {Baydin}, \citenamefont {Pearlmutter}, \citenamefont {Syme}, \citenamefont {Wood},\ and\ \citenamefont {Torr}}]{Baydin2022}%
  \BibitemOpen
  \bibfield  {author} {\bibinfo {author} {\bibfnamefont {A.~G.}\ \bibnamefont {Baydin}}, \bibinfo {author} {\bibfnamefont {B.~A.}\ \bibnamefont {Pearlmutter}}, \bibinfo {author} {\bibfnamefont {D.}~\bibnamefont {Syme}}, \bibinfo {author} {\bibfnamefont {F.}~\bibnamefont {Wood}},\ and\ \bibinfo {author} {\bibfnamefont {P.}~\bibnamefont {Torr}},\ }\bibfield  {title} {\bibinfo {title} {Gradients without backpropagation},\ }\href {https://doi.org/https://doi.org/10.48550/arXiv.2202.08587} {\bibfield  {journal} {\bibinfo  {journal} {ArXiv:2202.08587 [cs]}\ } (\bibinfo {year} {2022})}\BibitemShut {NoStop}%
\bibitem [{\citenamefont {Ren}\ \emph {et~al.}(2023)\citenamefont {Ren}, \citenamefont {Kornblith}, \citenamefont {Liao},\ and\ \citenamefont {Hinton}}]{Ren2022}%
  \BibitemOpen
  \bibfield  {author} {\bibinfo {author} {\bibfnamefont {M.}~\bibnamefont {Ren}}, \bibinfo {author} {\bibfnamefont {S.}~\bibnamefont {Kornblith}}, \bibinfo {author} {\bibfnamefont {R.}~\bibnamefont {Liao}},\ and\ \bibinfo {author} {\bibfnamefont {G.}~\bibnamefont {Hinton}},\ }\bibfield  {title} {\bibinfo {title} {Scaling forward gradient with local losses},\ }in\ \href {https://openreview.net/forum?id=JxpBP1JM15-} {\emph {\bibinfo {booktitle} {ICLR}}},\ Vol.~\bibinfo {volume} {11}\ (\bibinfo {year} {2023})\BibitemShut {NoStop}%
\bibitem [{\citenamefont {Ahmad}\ \emph {et~al.}(2023)\citenamefont {Ahmad}, \citenamefont {Schrader},\ and\ \citenamefont {van Gerven}}]{Ahmad2022}%
  \BibitemOpen
  \bibfield  {author} {\bibinfo {author} {\bibfnamefont {N.}~\bibnamefont {Ahmad}}, \bibinfo {author} {\bibfnamefont {E.}~\bibnamefont {Schrader}},\ and\ \bibinfo {author} {\bibfnamefont {M.}~\bibnamefont {van Gerven}},\ }\bibfield  {title} {\bibinfo {title} {Constrained parameter inference as a principle for learning},\ }\href {https://openreview.net/forum?id=CUDdbTT1QC} {\bibfield  {journal} {\bibinfo  {journal} {Trans. Mach. Learn. Res.}\ } (\bibinfo {year} {2023})}\BibitemShut {NoStop}%
\bibitem [{\citenamefont {Schnitzspan}\ \emph {et~al.}(2023{\natexlab{a}})\citenamefont {Schnitzspan}, \citenamefont {Kl\"aui},\ and\ \citenamefont {Jakob}}]{Schnitzspan2023}%
  \BibitemOpen
  \bibfield  {author} {\bibinfo {author} {\bibfnamefont {L.}~\bibnamefont {Schnitzspan}}, \bibinfo {author} {\bibfnamefont {M.}~\bibnamefont {Kl\"aui}},\ and\ \bibinfo {author} {\bibfnamefont {G.}~\bibnamefont {Jakob}},\ }\bibfield  {title} {\bibinfo {title} {Nanosecond true-random-number generation with superparamagnetic tunnel junctions: Identification of joule heating and spin-transfer-torque effects},\ }\href {https://doi.org/10.1103/PhysRevApplied.20.024002} {\bibfield  {journal} {\bibinfo  {journal} {Phys. Rev. Appl.}\ }\textbf {\bibinfo {volume} {20}},\ \bibinfo {pages} {024002} (\bibinfo {year} {2023}{\natexlab{a}})}\BibitemShut {NoStop}%
\bibitem [{\citenamefont {Hayakawa}\ \emph {et~al.}(2021)\citenamefont {Hayakawa}, \citenamefont {Kanai}, \citenamefont {Funatsu}, \citenamefont {Igarashi}, \citenamefont {Jinnai}, \citenamefont {Borders}, \citenamefont {Ohno},\ and\ \citenamefont {Fukami}}]{Hayakawa2021}%
  \BibitemOpen
  \bibfield  {author} {\bibinfo {author} {\bibfnamefont {K.}~\bibnamefont {Hayakawa}}, \bibinfo {author} {\bibfnamefont {S.}~\bibnamefont {Kanai}}, \bibinfo {author} {\bibfnamefont {T.}~\bibnamefont {Funatsu}}, \bibinfo {author} {\bibfnamefont {J.}~\bibnamefont {Igarashi}}, \bibinfo {author} {\bibfnamefont {B.}~\bibnamefont {Jinnai}}, \bibinfo {author} {\bibfnamefont {W.}~\bibnamefont {Borders}}, \bibinfo {author} {\bibfnamefont {H.}~\bibnamefont {Ohno}},\ and\ \bibinfo {author} {\bibfnamefont {S.}~\bibnamefont {Fukami}},\ }\bibfield  {title} {\bibinfo {title} {Nanosecond random telegraph noise in in-plane magnetic tunnel junctions},\ }\href {https://doi.org/10.1103/PhysRevLett.126.117202} {\bibfield  {journal} {\bibinfo  {journal} {Phys. Rev. Lett.}\ }\textbf {\bibinfo {volume} {126}},\ \bibinfo {pages} {117202} (\bibinfo {year} {2021})}\BibitemShut {NoStop}%
\bibitem [{\citenamefont {Safranski}\ \emph {et~al.}(2021)\citenamefont {Safranski}, \citenamefont {Kaiser}, \citenamefont {Trouilloud}, \citenamefont {Hashemi}, \citenamefont {Hu},\ and\ \citenamefont {Sun}}]{Safranski2021}%
  \BibitemOpen
  \bibfield  {author} {\bibinfo {author} {\bibfnamefont {C.}~\bibnamefont {Safranski}}, \bibinfo {author} {\bibfnamefont {J.}~\bibnamefont {Kaiser}}, \bibinfo {author} {\bibfnamefont {P.}~\bibnamefont {Trouilloud}}, \bibinfo {author} {\bibfnamefont {P.}~\bibnamefont {Hashemi}}, \bibinfo {author} {\bibfnamefont {G.}~\bibnamefont {Hu}},\ and\ \bibinfo {author} {\bibfnamefont {J.~Z.}\ \bibnamefont {Sun}},\ }\bibfield  {title} {\bibinfo {title} {Demonstration of nanosecond operation in stochastic magnetic tunnel junctions},\ }\href {https://doi.org/10.1021/acs.nanolett.0c04652} {\bibfield  {journal} {\bibinfo  {journal} {Nano Lett.}\ }\textbf {\bibinfo {volume} {21}},\ \bibinfo {pages} {2040} (\bibinfo {year} {2021})}\BibitemShut {NoStop}%
\bibitem [{\citenamefont {N{\'e}el}(1949)}]{Neel1949}%
  \BibitemOpen
  \bibfield  {author} {\bibinfo {author} {\bibfnamefont {L.}~\bibnamefont {N{\'e}el}},\ }\bibfield  {title} {\bibinfo {title} {Th{\'e}orie du tra{\^\i}nage magn{\'e}tique des ferromagn{\'e}tiques en grains fins avec application aux terres cuites},\ }in\ \href@noop {} {\emph {\bibinfo {booktitle} {Ann. Geophys.}}},\ Vol.~\bibinfo {volume} {5}\ (\bibinfo {year} {1949})\ pp.\ \bibinfo {pages} {99--136}\BibitemShut {NoStop}%
\bibitem [{\citenamefont {Schnitzspan}\ \emph {et~al.}(2020)\citenamefont {Schnitzspan}, \citenamefont {Cramer}, \citenamefont {Kubik}, \citenamefont {Tarequzzaman}, \citenamefont {Jakob},\ and\ \citenamefont {Kl{\"a}ui}}]{Schnitzspan2020}%
  \BibitemOpen
  \bibfield  {author} {\bibinfo {author} {\bibfnamefont {L.}~\bibnamefont {Schnitzspan}}, \bibinfo {author} {\bibfnamefont {J.}~\bibnamefont {Cramer}}, \bibinfo {author} {\bibfnamefont {J.}~\bibnamefont {Kubik}}, \bibinfo {author} {\bibfnamefont {M.}~\bibnamefont {Tarequzzaman}}, \bibinfo {author} {\bibfnamefont {G.}~\bibnamefont {Jakob}},\ and\ \bibinfo {author} {\bibfnamefont {M.}~\bibnamefont {Kl{\"a}ui}},\ }\bibfield  {title} {\bibinfo {title} {Impact of annealing temperature on tunneling magnetoresistance multilayer stacks},\ }\href {https://doi.org/10.1109/LMAG.2020.3005381} {\bibfield  {journal} {\bibinfo  {journal} {IEEE Magn. Lett.}\ }\textbf {\bibinfo {volume} {11}},\ \bibinfo {pages} {1} (\bibinfo {year} {2020})}\BibitemShut {NoStop}%
\bibitem [{\citenamefont {Schnitzspan}\ \emph {et~al.}(2023{\natexlab{b}})\citenamefont {Schnitzspan}, \citenamefont {Kläui},\ and\ \citenamefont {Jakob}}]{Schnitzspan2023b}%
  \BibitemOpen
  \bibfield  {author} {\bibinfo {author} {\bibfnamefont {L.}~\bibnamefont {Schnitzspan}}, \bibinfo {author} {\bibfnamefont {M.}~\bibnamefont {Kläui}},\ and\ \bibinfo {author} {\bibfnamefont {G.}~\bibnamefont {Jakob}},\ }\bibfield  {title} {\bibinfo {title} {Electrical coupling of superparamagnetic tunnel junctions mediated by spin-transfer-torques},\ }\href {https://doi.org/10.1063/5.0169679/2925723} {\bibfield  {journal} {\bibinfo  {journal} {Appl. Phys. Lett.}\ }\textbf {\bibinfo {volume} {123}} (\bibinfo {year} {2023}{\natexlab{b}})}\BibitemShut {NoStop}%
\bibitem [{\citenamefont {Fernández}\ \emph {et~al.}(2024)\citenamefont {Fernández}, \citenamefont {Keemink},\ and\ \citenamefont {van Gerven}}]{fernandez2024}%
  \BibitemOpen
  \bibfield  {author} {\bibinfo {author} {\bibfnamefont {J.~G.}\ \bibnamefont {Fernández}}, \bibinfo {author} {\bibfnamefont {S.}~\bibnamefont {Keemink}},\ and\ \bibinfo {author} {\bibfnamefont {M.}~\bibnamefont {van Gerven}},\ }\bibfield  {title} {\bibinfo {title} {Gradient-free training of recurrent neural networks using random perturbations},\ }\href {https://doi.org/10.3389/fnins.2024.1439155} {\bibfield  {journal} {\bibinfo  {journal} {Front. Neurosci.}\ }\textbf {\bibinfo {volume} {18}} (\bibinfo {year} {2024})}\BibitemShut {NoStop}%
\bibitem [{\citenamefont {Yorinks}()}]{TelemetrixArduino}%
  \BibitemOpen
  \bibfield  {author} {\bibinfo {author} {\bibfnamefont {A.}~\bibnamefont {Yorinks}},\ }\href@noop {} {\bibinfo {title} {The telemetrix project}},\ \bibinfo {howpublished} {\url{https://mryslab.github.io/telemetrix}},\ \bibinfo {note} {accessed: 2010-09-30}\BibitemShut {NoStop}%
\bibitem [{\citenamefont {Belouze}(2022)}]{Belouze2022}%
  \BibitemOpen
  \bibfield  {author} {\bibinfo {author} {\bibfnamefont {G.}~\bibnamefont {Belouze}},\ }\bibfield  {title} {\bibinfo {title} {Optimization without backpropagation},\ }\href {https://doi.org/10.48550/arXiv.2209.06302} {\bibfield  {journal} {\bibinfo  {journal} {ArXiv:2209.06302v1 [cs.LG]}\ } (\bibinfo {year} {2022})}\BibitemShut {NoStop}%
\bibitem [{\citenamefont {Jaeger}\ \emph {et~al.}(2023)\citenamefont {Jaeger}, \citenamefont {Noheda},\ and\ \citenamefont {van~der Wiel}}]{Jaeger2023}%
  \BibitemOpen
  \bibfield  {author} {\bibinfo {author} {\bibfnamefont {H.}~\bibnamefont {Jaeger}}, \bibinfo {author} {\bibfnamefont {B.}~\bibnamefont {Noheda}},\ and\ \bibinfo {author} {\bibfnamefont {W.~G.}\ \bibnamefont {van~der Wiel}},\ }\bibfield  {title} {\bibinfo {title} {Toward a formal theory for computing machines made out of whatever physics offers},\ }\href {https://doi.org/10.1038/s41467-023-40533-1} {\bibfield  {journal} {\bibinfo  {journal} {Nat. Commun.}\ }\textbf {\bibinfo {volume} {14}},\ \bibinfo {pages} {1} (\bibinfo {year} {2023})}\BibitemShut {NoStop}%
\bibitem [{\citenamefont {Kammerbauer}\ and\ \citenamefont {Schnitzspan}(2025)}]{sMTJ2025}%
  \BibitemOpen
  \bibfield  {author} {\bibinfo {author} {\bibfnamefont {F.}~\bibnamefont {Kammerbauer}}\ and\ \bibinfo {author} {\bibfnamefont {L.}~\bibnamefont {Schnitzspan}},\ }\bibfield  {title} {\bibinfo {title} {Time series s{MTJ} switching data},\ }\href {https://doi.org/10.5281/zenodo.15222667} {10.5281/zenodo.15222667} (\bibinfo {year} {2025})\BibitemShut {NoStop}%
\end{thebibliography}%
\end{document}